\documentclass[12pt]{article}

\usepackage{latexsym}
\usepackage{amsmath}
\usepackage{amsfonts}
\usepackage{mathrsfs}
\usepackage{dsfont}
\usepackage{multibox}
\usepackage{verbatim}

\usepackage[colorlinks=true,linkcolor=black,citecolor=black,urlcolor=black,filecolor=black]{hyperref}

 \csname
@addtoreset\endcsname{equation}{section}

\def\b{\beta}

\def\d{\delta}

\def\e{\epsilon}

\def\h{\eta}

\def\l{\lambda}
\def\L{\Lambda}
\def\m{\mu}
\def\n{\nu}
\def\x{\xi}

\def\p{\pi}

\def\vf{\varphi}

\def\o{\omega}

\def\cD{{\cal D}}

\def\cW{{\cal W}}

\def\be{\begin{equation}}
\def\ee{\end{equation}}
\def\bea{\begin{eqnarray}}
\def\eea{\end{eqnarray}}
\def\ba{\begin{array}}
\def\ea{\end{array}}

\def\tr{\text{tr}}

\def\ww{\wedge}

\def\12{\frac{1}{2}}
\def\pr{\partial}

\begin{document}

\begin{flushright}
AEI-2011-074
\end{flushright}

\vspace{1pt}

\begin{center}


{\Large\bfseries Higher Spins in $D=2+1$}\\


\vspace{18pt}
{\sc Andrea Campoleoni}
\vspace{7pt}

{\sl\small
Max-Planck-Institut f{\"u}r Gravitationsphysik\\
Albert-Einstein-Institut\\
Am M{\"u}hlenberg 1\\
14476 Golm,\ GERMANY\\[7pt]

{\it andrea.campoleoni@aei.mpg.de}} 

\vspace{14pt}

\end{center}

\begin{abstract}
\noindent
We give a brief overview of some three-dimensional toy models for higher-spin interactions. We first review the construction of pure higher-spin gauge theories in terms of Chern-Simons theories. We then discuss how this setup could be modified along the lines of the known topologically massive theories.

\vspace{10pt}
\noindent
\emph{Based on the talk presented at the 49th Course of the International School of Subnuclear Physics, Erice, Italy, 24 June -- 3 July 2011}
\end{abstract}

\section{Introduction}\label{sec:intro}

The study of field theories for particles of arbitrary spin has a long and variegated history,\footnote{For a more complete overview on higher-spin field theories see, for instance, \cite{reviews}.} but in the last years most of the efforts focussed on gauge theories, to wit on massless particles. The motivations come from String Theory -- whose higher-spin excitations could acquire their masses via the breaking of a ``hidden'' gauge symmetry (see e.g.\ \cite{vertices-string} and refs therein) -- and, more recently, from the study of holographic dualities \cite{AdS-CFT}.
We are still far from having a full control on these proposals but -- in spite of various renowned no-go theorems (see e.g.\ \cite{no-go} for a review) -- we now have examples of consistent interactions to analyse. The old negative results thus only signal that higher-spin interactions are ``unconventional'', mainly on account of the higher-derivative nature of the couplings \cite{vertices,vertices-string}. In flat space the construction of a complete higher-spin gauge theory is still ongoing \cite{nonlocal}, but in the presence of a cosmological constant we already have at our disposal the Vasiliev equations \cite{Vasiliev}, which are a set of non-linear field equations for an infinite tower of gauge fields of increasing spin. The need for a cosmological constant, however, makes it difficult to compare Vasiliev's models with String Theory. On the other hand, it suggested to consider them in the AdS/CFT context \cite{AdS-CFT}. In this case the comparison with the candidate boundary duals has to face the ``unconventional'' formulation of the Vasiliev equations, and thus often requires the development of ad hoc techniques (see e.g.\ \cite{checks-HD} and refs therein).

These open questions call for models where one can address at least some of them in a simplified setup. Various low-dimensional field theories fit in this scheme even if, at a first glance, they seem to have no chance to account for higher spins. For instance, in a space-time of dimension $D=2+1$ the Poincar\'e group does not admit massless representations of arbitrary spin \cite{Binegar}. Nevertheless, one can still consider the Fronsdal equation\footnote{Here and in the following a couple of parentheses (brackets) denotes the (anti)symmetrisation of the indices it encloses, with the minimum number of terms and without any normalisation factor. Repeated indices denote a summation, and we adopt the mostly plus convention for the space-time metric.}
\be \label{fronsdal}
\Box\, \vf_{\m_1 \ldots\, \m_s} \,-\, \pr_{(\m_1} \pr^\l \vf_{\m_2 \ldots\, \m_s)\l} \,+\, \pr_{(\m_1}\pr_{\m_2}\vf_{\m_3 \ldots\, \m_s)\l}{}^\l \,=\, 0 \, ,
\ee
whose solutions describe, for $D > 3$, the free propagation of a massless spin-$s$ particle \cite{Fronsdal}. Here $\vf_{\m_1 \ldots\, \m_s}$ is a symmetric tensor of rank $s$, and eq.~\eqref{fronsdal} is left invariant by the gauge transformations
\be \label{gauge}
\d \vf_{\m_1 \ldots\, \m_s} =\, \pr_{(\m_1} \x_{\m_2 \ldots\, \m_s)} \qquad \textrm{with} \quad \x_{\m_1 \ldots\, \m_{s-3}\l}{}^\l \,=\, 0 \, ,
\ee
that are crucial to ensure the decoupling of unphysical polarisations. If one wants to derive \eqref{fronsdal} from a Lagrangian, one also has to force the double trace of $\vf_{\m_1 \ldots\, \m_s}$ to vanish, either directly or by adding a Lagrange multiplier (see e.g. \cite{reviews}). One of the main goals of the current research on higher spins is the classification of the consistent non-linear deformations of \eqref{fronsdal} and \eqref{gauge}, and their comparison with Vasiliev's models.\footnote{We focussed on \eqref{fronsdal} for simplicity, but for $D>5$ the description of all irreps of the Poincar\'e group requires fields with multiple groups of symmetrised indices (see e.g. \cite{mixed}).} One can pursue both tasks even in $D=2+1$: the price to pay is that \eqref{fronsdal} does not propagate local degrees of freedom when $s>1$, while the advantage is the opportunity to deal with questions that are beyond reach in higher space-time dimensions. 

The possible three-dimensional deformations include a class of models that is particularly appealing for its simplicity: the full non-linear theory is described by a Chern-Simons action \cite{Blencowe} (see also \cite{spin3}). In sect.~\ref{sec:gauge} we review how to select the relevant Chern-Simons actions and in sect.~\ref{sec:top} we discuss how one could extend this setup by adding topological mass terms. Along the way we briefly recall some recent applications of these models.

\section{Chern-Simons theories}\label{sec:gauge}

For $s=2$ the Fronsdal equation \eqref{fronsdal} imposes the vanishing of the linearised Ricci tensor. The study of its deformations thus aims at understanding higher-spin interactions along the lines of the metric formulation of gravity. However, alternative descriptions of the free dynamics are available \cite{reviews}. For instance, one can mimic the frame formulation of gravity and associate to each particle a couple of differential forms \cite{frame}. The second-order Fronsdal action for the symmetric field $\vf_{\m_1 \ldots\, \m_s}$ is replaced with a first-order action for a vielbein-like field 
\be \label{vielbein}
e^{\,a_1 \ldots\, a_{s-1}} \,=\, e_\m{}^{a_1 \ldots\, a_{s-1}}\, dx^\m \, ,
\ee
and an auxiliary field
\be \label{spinconn}
\o^{\,a_1 \ldots\, a_{s-1}\,,\,b} \,=\, \o_\m{}^{a_1 \ldots\, a_{s-1}\,,\,b}\, dx^\m \, ,
\ee
which generalises the spin connection. The vielbein-like field is traceless and fully symmetric in the fiber indices, while the auxiliary field is traceless and satisfies the condition $\o^{(a_1 \ldots a_{s-1}\,,\,b)} \,=\, 0$. For more details we refer to \cite{spin3}. For the purpose of this note it is sufficient to recall that the free action is invariant under the transformations
\begin{subequations} \label{frame_gauge}
\begin{align}
& \d\, e_\m{}^{a_1 \ldots\, a_{s-1}} \, = \, \pr_\m \, \x^{\,a_1 \ldots\, a_{s-1}} \, + \, \bar{e}_{\m\,,\,b}\, \L^{a_1 \ldots\, a_{s-1}\,,\,b} \, , \label{frame_gauge_viel} \\[5pt]
& \d\, \o_\m{}^{a_1 \ldots\, a_{s-1}\,,\,b} \, = \, \pr_\m \, \L^{a_1 \ldots\, a_{s-1}\,,\,b} \, ,
\end{align}
\end{subequations}
where $\bar{e}_\m{}^a$ denotes the background vielbein. The gauge transformations generated by $\L$ allow one to recover the Fronsdal field by gauging away the non-symmetric component of the vielbein. Those generated by $\x$ then reproduce the gauge transformations \eqref{gauge} (see e.g. \cite{spin3}). Also in this context one can classify the non-linear deformations of the free action that admit a gauge symmetry reducing to \eqref{frame_gauge} at the linearised level. 

In $D=2+1$ there is a simple class of deformations that is not available in higher space-time dimensions \cite{Blencowe}, and all ingredients to build it are already visible in the spin-two case. To spot them, it is convenient to combine dreibein and spin connection in a one-form taking values in the algebra of isometries of the maximally symmetric solution of the field equations,
\be
A \,=\, e \,+\, \o := \left(\, e_\m{}^a P_a \,+\, \frac{1}{2}\ \o_\m{}^{a,b}\, M_{ab} \,\right) dx^\m\, .
\ee
Here the $M_{ab}$ denote the generators of the Lorentz group, while the $P_a$ denote the translation generators:
\be \label{algebra}
\begin{split}
& [\,P_a\,,\,P_b\,]=\frac{1}{l^2}\, M_{ab} \, , \qquad [\,P_a\,,\,M_{bc}\,] = \eta_{a[b}P_{c]} \, ,\\
& [\, M_{ab} \,,\, M_{cd}\,] = \h_{a[c}M_{d]b} \,-\, \h_{b[c}M_{d]a} \, .
\end{split}
\ee
The factor $l$ is related to the cosmological constant by $l = 1/\sqrt{-\,\L}$, so that \eqref{algebra} gives the Poincar\'e algebra for $l \to \infty$ and either the $so(2,D-1)$ or $so(1,D)$ algebras for positive or negative values of $l^2$. A Poincar\'e -- or (Anti) de Sitter -- gauge transformation $\d A = d\l + [A,\l]$ then induces
\begin{subequations} \label{gauge_fields}
\begin{align}
& \d e \,=\, d\x \,+\, [\, \o \,,\, \x \,] \,+\, [\, e \,,\, \L \,]  \,=\, \cD\x \,+\, [\, e \,,\, \L \,] \, , \label{gauge_viel} \\
& \d \o \,=\, d\L \,+\, [\, \o \,,\, \L \,] \,+\, [\, e \,,\, \x \,] \,=\, \cD\L \,+\, [\, e \,,\, \x \,] \, , \label{gauge_spin}
\end{align}
\end{subequations}
where we split the gauge parameter as
\be
\l \,=\, \x \,+\, \L :=\, \x^a P_a \,+\, \frac{1}{2}\, \L^{a,b} M_{ab} \, ,
\ee
and we denoted by $\cD$ the covariant exterior derivative. At the linearised level \eqref{gauge_fields} reduces to \eqref{frame_gauge} or to its (A)dS counterpart. However, in a space-time of generic dimension the Einstein-Hilbert action is \emph{not} invariant under the full set of transformations \eqref{gauge_fields}, but only under local Lorentz transformations. The non-linear deformation of \eqref{frame_gauge_viel} leads to diffeomorphisms rather than local translations. 

On the other hand, in $D=2+1$ the Einstein-Hilbert action reads
\be \label{EH}
S_{EH} =\, \frac{1}{16\pi G} \int \e_{abc} \left(\, e^a \ww R^{bc} \,+\, \frac{1}{3\,l^2}\, e^a \ww e^b \ww e^c \,\right) \, ,
\ee
where we introduced the Riemann curvature
\be \label{riemann}
R :=\, d\o + \o \ww \o \, .
\ee
Up to boundary terms, the variation of \eqref{EH} under \eqref{gauge_fields} is
\be \label{var_EH}
\d S_{EH} \,=\, - \, \frac{1}{16\pi G} \int \e_{abc} \left(\, \x^a\, \cD R^{bc} \,\right) \,=\, 0 \, .
\ee
The remainder vanishes on account of the Bianchi identity, but setting the dimension of space-time to $D=2+1$ is instrumental in obtaining \eqref{var_EH}. For instance, the variation of the Einstein-Hilbert action in $D=3+1$ is
\be
\d S_{EH} \,\sim \int \e_{abcd}\, \cD\x^a \ww e^b \ww  R^{cd} \,= \,-\, \int \e_{abcd}\, \x^a\, T^b \ww  R^{cd} \,\neq\, 0 \, ,
\ee
and does not vanish off shell due to the presence of the torsion two-form,
\be \label{torsion}
T :=\, \cD e \,=\, de \,+ \left(\, \o \ww e + e \ww \o \,\right) \, .
\ee

As a result, \emph{only} in three dimensions Einstein gravity is a gauge theory for the full group of isometries of the vacuum. Moreover, the field equations following from \eqref{EH} are
\be \label{eq}
R \,+\, \frac{1}{l^2}\, e \ww e \,=\, 0 \, , \qquad\qquad 
T \,=\, 0 \, ,
\ee
and imply that the curvature of $A$ vanishes. One can thus conclude that, up to boundary terms, \eqref{EH} should be equivalent to a Chern-Simons action with gauge algebra \eqref{algebra}. This was indeed realised more than twenty years ago in \cite{Witten}, where \eqref{EH} was recovered from 
\be\label{CS}
S_{CS} =\, \frac{k}{4\pi} \int \tr \left(\, A \ww dA \,+\, \frac{2}{3}\, A \ww A \ww A \,\right)  \quad \textrm{with} \ k \,=\, \frac{1}{4G} \ ,
\ee 
using the following bilinear invariant form of \eqref{algebra}:\footnote{This invariant form exists only in $D=2+1$, where \eqref{algebra} is not simple even for finite values of $l$ \cite{Witten}.}
\be \label{trace}
\tr\left(P_aP_b\right) \,=\, 0 \, , \quad \tr\left(P_a M_{bc}\right) \,=\, \e_{abc} \, , \quad \tr\left(M_{ab}M_{cd}\right) \,=\, 0 \, . 
\ee
The invariance of \eqref{EH} under local translations (i.e.\ the transformations generated by $\x$ in \eqref{gauge_fields}) does not imply that its gauge group is bigger than expected. In fact, \emph{on shell} a local translation generated by $\x^a = v^\m e_\m{}^{a}$ is equivalent to the Lie derivative along $v^\m$ \cite{Witten}. This is the key to three-dimensional simplicity: one can trade diffeomorphisms for gauge translations! In the higher-spin context it is not yet clear what the precise analogue of diffeomorphism invariance is, but it is relatively straightforward to extend the Chern-Simons formulation of the dynamics \cite{Blencowe}.

To this end, one should notice that in $D=2+1$ the higher-spin vielbein and spin connection \eqref{vielbein} and \eqref{spinconn} have the same number of independent components. For instance, in the $s=2$ case this can be made manifest by
\be
\o^a \,=\, \frac{1}{2}\, \e^{abc}\, \o_{bc} \quad \leftrightarrow \quad M_a \,=\, -\, \frac{1}{2}\, \e_{abc}\, M^{bc} \, ,
\ee
that induces a rewriting of \eqref{algebra} as
\be \label{algebra2}
[\,P_a\,,\,P_b\,] = \frac{1}{l^2}\,\e_{abc}\,M^c , \quad [\,M_a\,,\,P_b\,] = \e_{abc}\,P^c , \quad [\,M_a\,,\,M_b\,] = \e_{abc}\,M^c  .
\ee
The possibility to deal with vielbeins and spin connections with the same structure suggests to build higher-spin gauge theories out of Chern-Simons theories with a non-compact gauge algebra of the form
\begin{subequations} \label{algebra_big}
\begin{align}
[\, P_A \,,\, P_B \,] \,& =\, \frac{1}{l^2}\, f_{AB}{}^C\, M_C \,, \\
[\, M_A \,,\, P_B \,] \,& =\, f_{AB}{}^C\, P_C \,, \label{big-PM} \\[4pt]
[\, M_A \,,\, M_B \,] \,& =\, f_{AB}{}^C\, M_C \,. \label{big-lorentz}
\end{align}
\end{subequations}
The structure constants that appear in the three sets of commutators are identical, and thus there is an equal number of $P_A$ and $M_A$ to be contracted with \eqref{vielbein} and \eqref{spinconn}. Any $sl(2,\mathbb{R})$ subalgebra of \eqref{big-lorentz} can then play the role of the Lorentz algebra since $so(1,2) \sim sl(2,\mathbb{R})$. Once one chooses it, the adjoint action of the Lorentz generators $M_a \subset M_A$ branches the $P_A$ and $M_A$ into identical $sl(2,\mathbb{R})$-irreducible subsets.\footnote{This is the crucial condition that one has to impose on the gauge algebra. One can generalise \eqref{algebra_big} provided the two sets of $P_A$ and $M_A$ are still branched into identical $sl(2,\mathbb{R})$-irreducible subsets. An example can be found in sect.~2.3 of \cite{spin3}.} Each of them has dimension $2s+1$ with $2s \in \mathbb{N}$ and, for integer $s$, this number corresponds to the number of independent components of a traceless symmetric tensor of rank $s$. This suffices to identify the corresponding vielbeins and spin connections with \eqref{vielbein} and \eqref{spinconn}. In the linearised regime \eqref{gauge_fields} then induces the transformations \eqref{frame_gauge} on all fields emerging from this branching \cite{spin3} (see also \cite{Wlambda} for the treatment of half-integer values of $s$). This completes the argument that any Chern-Simons action based on an algebra of the form \eqref{algebra_big} gives a non-linear deformation of the free actions of a proper set of higher-spin gauge fields.

A key feature of \eqref{algebra_big} is the enlargement of the Lorentz algebra to \eqref{big-lorentz}. A field $\vf_{\m_1 \ldots\, \m_s}$ must be invariant under Lorentz-like transformations since they do not have any counterpart in the Fronsdal approach. In the case of the metric, the only combination of the vielbeins that is invariant under generalised Lorentz transformations is
\be \label{metric}
g_{\m\n} \,\sim\, l^2\, \tr\left(e_\m e_\n\right) \,=\, l^2\, e_\m{}^A e_\n{}^B\, \tr\left(P_A P_B \right) \, .
\ee
Here $e$ is the full higher-spin vielbein $e^A P_A$ \cite{spin3}, so that the metric depends on the higher-spin vielbeins! Unfortunately, requiring Lorentz-like invariance does not suffice to fix the form of higher-rank fields, but additional results can be found in \cite{Wlambda}. In spite of this limitation, \eqref{metric} already played a role in the study of BTZ-like solutions with higher-spin hairs and of the effects of higher-spin gauge transformations on the causal structure of space-time \cite{bh1}.

To conclude, let us recall that for $l^2 > 0$ there is a simple way to satisfy \eqref{algebra_big}. It is sufficient to consider algebras of the form $\mathfrak{g} \oplus \mathfrak{g}$ and to define the $P_A$ and $M_A$ as
\be \label{gen}
P_A \,=\, \frac{1}{l} \left(\, T_A \,-\, \widetilde{T}_A \,\right) \, , \qquad\quad M_A \,=\, T_A \,+\, \tilde{T}_A \, ,
\ee
where $T_A$ and $\widetilde{T}_A$ denote the generators of the two copies of $\mathfrak{g}$. This setup was used in \cite{HR,spin3} to show that the asymptotic symmetries of three-dimensional higher-spin gauge theories on asymptotically (A)dS backgrounds are given by non-linear $\cW$-algebras (see also \cite{checkW,Wlambda}). In the previous discussion nothing prevents the algebras $\mathfrak{g}$ from having a finite dimension, so that in $D=2+1$ one can ``truncate'' the spectra of higher-spin gauge theories. An example of this type is given by $\mathfrak{g} \equiv sl(N,\mathbb{R})$, where the Lorentz algebra is identified with the principal embedding of $sl(2,\mathbb{R})$ in the diagonal $sl(N,\mathbb{R})$ generated by the $M_A$. In this case, the Chern-Simons theory describes the interactions of a tower of symmetric tensors of ranks $2,3,\ldots,N$ (see also \cite{exclusion} for a discussion of the quantum spectrum). On the other hand, it is possible to consider infinite-dimensional algebras $\mathfrak{g}$ so as to mimic the gauge algebras underlying Vasiliev's construction. In this context, Gaberdiel and Gopakumar proposed a holographic duality between Vasiliev's models in $D=2+1$ \cite{vas-prok} and suitable limits of minimal model CFT's with $\cW_N$ symmetry \cite{conjecture} (see also \cite{checks}).

The construction \eqref{gen} also allows an interesting rewriting of the action \eqref{CS}. A generic bilinear invariant form on $\mathfrak{g} \oplus \mathfrak{g}$ can be cast in the form
\be
\tr\, (T_AT_B) \,=\, c_1\, \h_{AB} \, , \quad \tr\, (T_A\widetilde{T}_B) \,=\, 0 \, , \quad \tr\, (\widetilde{T}_A\widetilde{T}_B) \,=\, c_2\, \h_{AB} \, ,
\ee
so that \eqref{CS} can be rewritten as
\be
S_{\mathfrak{g} \oplus \mathfrak{g}} \,=\, c_1\, S_{CS}[A] \,+\, c_2\, S_{CS}[\widetilde{A}] \, ,
\ee
where now both $A$ and $\widetilde{A}$ take values in a single copy of $\mathfrak{g}$: $A = A^BT_B$ and $\widetilde{A} = \widetilde{A}^B T_B$. If one defines
\be
e \,=\, \frac{l}{2} \left( A - \widetilde{A} \,\right) \, , \qquad \o \,=\, \frac{1}{2} \left( A + \widetilde{A} \,\right) \, ,
\ee
and chooses $c_1 = l$ and $c_2 = - \, l$, one then obtains
\be \label{EH_big}
S_{\mathfrak{g} \oplus \mathfrak{g}} \,=\, \frac{1}{8\pi G} \int \tr\left(\, e \ww R \,+\, \frac{1}{3l^2}\, e \ww e \ww e  \,\right) \, ,
\ee
where $R$ is defined as in \eqref{riemann}, all fields take values in a single copy of $\mathfrak{g}$ and the trace is defined by $\tr\left(T_AT_B\right)=\eta_{AB}$. This is the direct generalisation of \eqref{EH} for a generic higher-spin gauge theory. In this respect, the Chern-Simons form of the action is just a convenient tool to handle these models, but it is not crucial. One could also deal directly with \eqref{EH_big}, and this opens the way to the generalisations of \eqref{CS} that we shall briefly consider in the following section.

\section{Topologically massive theories}\label{sec:top}

The actions \eqref{EH_big} do not propagate local degrees of freedom, but in the $s=2$ case it is known how to make the graviton massive and propagating without breaking diffeomorphism invariance \cite{tmg}. In a first-order formalism this goal is achieved with the action \cite{1otmg}
\be \label{TMG}
S \,=\, S_{EH}\, + \, \frac{\m^{-1}}{16\p G} \int \tr\left( \o \ww d \o + \frac{2}{3} \, \o \ww \o \ww \o\right) + \int \tr\left(\, \b \ww T \,\right) ,
\ee
where we continue to follow the notation introduced in \eqref{EH_big}. Therefore, all fields are one-forms taking values in $sl(2,\mathbb{R}) \sim so(1,2)$. The one-form $\b$ is a Lagrange multiplier enforcing the torsion constraint, so that the elimination of all auxiliary fields would lead to the three-derivative action of \cite{tmg}.\footnote{Without the term $\b \ww T$ the action \eqref{TMG} is instead equivalent to \eqref{EH_big} \cite{Witten}. On the other hand, adding a term $e\ww T$ leads to a higher-spin generalisation of the Mielke-Baekler model \cite{MB}. The resulting action is still expressible as the difference of two Chern-Simons actions (by extending \cite{BV}), but the higher-spin torsion $T$ no longer vanishes on shell.} Moreover, \eqref{TMG} is manifestly invariant under diffeomorphisms and local Lorentz transformations.

It is thus natural to preserve the structure \eqref{TMG} and try and enlarge the tangent algebra $sl(2,\mathbb{R})$ to a generic semisimple $\mathfrak{g}$. However, the resulting action is still manifestly invariant under diffeomorphisms and extended Lorentz transformations of the form
\be \label{lor}
\d e \,=\, [\,e\,,\,\L\,] \, , \qquad \d \o \,=\, \cD\L \,, \qquad \d \b \,=\, [\,\b\,,\,\L\,] \, ,
\ee
but it is not possible to tune the gauge variation of $\b$ to cancel the variation of \eqref{TMG} under extended local translations of the form
\be \label{transl}
\d e \,=\, \cD\x \,, \qquad \d \o \,=\, \frac{1}{l^2}\, [\,e\,,\,\x\,] \,.
\ee
In both \eqref{lor} and \eqref{transl} fields and gauge parameters take values in $\mathfrak{g}$, so that we had to exhibit the $l$-dependence in \eqref{transl}. The gauge variation of \eqref{TMG} is 
\be \label{var_TMG}
\d S \,=\, \int \tr \left(\, \d\b \ww T \,+\, \b \ww [\,R \,+\, \frac{1}{l^2}\, e \ww e \,,\, \x \,] \,\right) \, .
\ee
At the linearised level $\d \b = \cD \x$ implies the vanishing of \eqref{var_TMG}: the remainder is proportional to the equations that are satisfied by the background dreibein and spin connection. At the non-linear level, however, one has at least to deform \eqref{transl} and $\d \b = \cD \x$ in order to preserve the linearised gauge symmetry. In the $s=2$ case we know how: the non-linear deformation leads to diffeomorphisms. In the higher-spin setup it is not clear whether \eqref{TMG} admits a larger symmetry group with respect to \emph{diff}$\times \mathfrak{g}$ and, possibly, what its precise structure is.

The enhancement of $sl(2,\mathbb{R})$ to $sl(N,\mathbb{R})$ was already discussed in \cite{tmhs1}, where the authors studied the linearised spectrum of the resulting theory. A similar analysis was performed in \cite{tmhs2} starting from the (A)dS generalisation of a quadratic action first proposed in \cite{DD}. Since this is the most general quadratic action with three derivatives displaying the Fronsdal gauge symmetry, the authors argued that it should be related to the linearisation of an action of the type \eqref{TMG}. A comparison of these results with a direct analysis of the constraint algebra of \eqref{TMG} would further clarify the nature of the symmetries of \eqref{TMG}, or open the way to possible completions preserving all gauge symmetries of \eqref{EH_big}.\footnote{An analysis of fermionic higher-spin topologically massive theories was performed in \cite{AD}, but the authors spotted some problems in coupling these theories to gravity.} The same result could be achieved by deforming the symmetries \eqref{transl} within the Noether procedure. 

Higher-order gauge invariant massive theories for higher spins were also considered in \cite{TMG2} in a metric-like formalism. In this case the linearised gauge transformations still have the form \eqref{gauge}, but the gauge parameter is not traceless. As a result, this proposal is not directly related to \eqref{EH_big}, that at the linearised level reproduces the Fronsdal action if one eliminates the spin connections via the generalised torsion constraint \cite{spin3}.

\vspace{15pt}
\noindent
{\bf Acknowledgments:}\hspace{5pt} I am grateful to S.~Fredenhagen, S.~Pfenninger and S.~Theisen for extensive discussions and collaboration on the material discussed in this note. I also would like to thank A.~Anabal\'on, X.~Bekaert, D.~Francia, E.~Joung, I.~V.~Melnikov, A.~Sagnotti, L.~Sindoni, M.~Sivakumar and M.~Taronna for stimulating discussions.


\end{document}